\documentclass[conference]{IEEEtran}
\IEEEoverridecommandlockouts
\usepackage{cite}
\usepackage{amsmath,amssymb,amsfonts}
\usepackage{algorithmic}
\usepackage{graphicx}
\usepackage{textcomp}
\usepackage{xcolor}
\usepackage{caption}
\usepackage{subcaption}
\usepackage{wrapfig}

\DeclareMathOperator*{\argmin}{arg\,min}

\begin{document}

\title{Vital Measurements of Hospitalized COVID-19 Patients as a Predictor of Long COVID: An EHR-based Cohort Study from the RECOVER Program in N3C}

\author{\IEEEauthorblockN{Sihang Jiang\IEEEauthorrefmark{1},
Johanna Loomba\IEEEauthorrefmark{2}, Suchetha Sharma\IEEEauthorrefmark{3},
Donald Brown\IEEEauthorrefmark{1}\IEEEauthorrefmark{3}\\
on behalf of the RECOVER consortium and the N3C consortium}

\IEEEauthorblockA{\IEEEauthorblockA{
\IEEEauthorrefmark{1}Department of Engineering of Systems and Environment, University of Virginia\\
\IEEEauthorrefmark{2}integrated Translational Health Research Institute of Virginia (iTHRIV), University of Virginia\\
\IEEEauthorrefmark{3}School of Data Science, University of Virginia\\
\IEEEauthorrefmark{1}\IEEEauthorrefmark{2}\IEEEauthorrefmark{3}Email: \{sj5yq, jjl4d, ss4jg, deb\}@virginia.edu
}
}
}
\makeatletter
\newcommand{\linebreakand}{%
  \end{@IEEEauthorhalign}
  \hfill\mbox{}\par
  \mbox{}\hfill\begin{@IEEEauthorhalign}
}
\makeatother

\makeatletter
\newcommand*{\rom}[1]{\expandafter\@slowromancap\romannumeral #1@}
\makeatother
\makeatletter
\def\ps@IEEEtitlepagestyle{%
  \def\@oddfoot{\mycopyrightnotice}%
  \def\@evenfoot{}%
}
\def\mycopyrightnotice{%
  {\footnotesize 978-1-6654-6819-0/22/\$31.00 ©2022 IEEE\hfill}
  \gdef\mycopyrightnotice{}
}

\maketitle

\begin{abstract}
It is shown that various symptoms could remain in the stage of post-acute sequelae of SARS-CoV-2 infection (PASC), otherwise known as Long COVID. A number of COVID patients suffer from heterogeneous symptoms, which severely impact recovery from the pandemic. While scientists are trying to give an unambiguous definition of Long COVID, efforts in prediction of Long COVID could play an important role in understanding the characteristic of this new disease. Vital measurements (e.g. oxygen saturation, heart rate, blood pressure) could reflect body's most basic functions and are measured regularly during hospitalization, so among patients diagnosed COVID positive and hospitalized, we analyze the vital measurements of first 7 days since the hospitalization start date to study the pattern of the vital measurements and predict Long COVID with the information from vital measurements.   
\end{abstract}


\section{Introduction}

Since the outbreak of COVID-19\footnote{The RECOVER's official terminology of COVID-19 is SARS-CoV-2.} pandemic in March 2020, numerous studies focus on the typical symptoms of COVID-19 patients and the characteristic of the transmission process. It is believed that typical symptoms of COVID-19 include fever, dry cough, and fatigue, often with pulmonary involvement, and the incubation period has an average of 5–7 days \cite{shi2020overview}. The term `Long COVID' is being used to describe the illness in people who have either recovered from COVID-19 but are still reporting lasting effects of the infection or have had the usual symptoms for far longer than would be expected \cite{mahase2020covid}. A research team from Italy studied 143 patients discharged from a Rome hospital after recovering from COVID-19, and found out that 87\% were experiencing at least one symptom after 60 days \cite{carfi2020persistent}. Common symptoms of Long COVID include profound fatigue, cough, breathlessness, muscle and body aches, chest heaviness or pressure and so on \cite{nabavi2020long}, while some patients reported difficulty doing daily activities, in addition to mental health issues \cite{Raveendran2021LongOverview}. 

The main risk factors for severe COVID-19 and hospital admission include older age, male sex, non-white ethnicity, disability, and pre-existing comorbidities \cite{crook2021long}. However, the risk factors of Long COVID are still generally unclear. Some risk factors of COVID-19 do not increase risk of Long COVID, such as male sex, obesity, diabetes, and cardiovascular disease; pre-existence of asthma has been found to be significantly associated with Long COVID \cite{sudre2021attributes}. Scientists have been trying to predict Long COVID with some of the risk factors. According to a recent study, for patients with a duration of COVID symptoms longer than 28 days, five symptoms during the first week that were most predictive were fatigue, headache, dyspnea, hoarse voice and myalgia \cite{sudre2021attributes}. Nevertheless, there is limit attention to the vital measurements of hospitalized COVID patients. Compared to symptoms, vital measurements are more frequently measured and available, making wonderful time series to reflect conditions of patients. What's more, oxygen saturation, heart rate and blood pressure are all quantitative variables, making it easy to be inputs of statistical machine learning models. Thus, exploring the pattern of vital measurements among hospitalized COVID patients in an early stage could help medical workers identify COVID patients with a high risk of Long COVID, give a better understanding of the new disease Long COVID, and make efforts to control the pandemic for social benefits.
\section{Related Work}
A recent study analyzed 4,182 incident cases of COVID-19 in which individuals are categorized as short, LC28, LC56 and intermediate \cite{sudre2021attributes}. This study applied random forest prediction models using personal characteristics and comorbidities, and the average AUC-ROC was 76.8\% in classifying between short COVID and LC28. Some strong predictors include increasing age and the number of symptoms during the first week. The National COVID Cohort Collaborative (N3C) \cite{haendel2021national} has collected abundant clinical data that can be used to understand the long term effects of COVID-19 and identify the clinical features of Long COVID \cite{Pfaff2021.10.18.21265168} \cite{pfaff2022identifying}. In this study, 924 features were selected from demographics, healthcare visits, medical conditions, and prescriptions. An XGBoost model was trained and tested on a set of 97,995 patients who had visited a long COVID clinic, and got an AUC-ROC of 92\%. Some of the important features include post-COVID outpatient utilisation, age, post-COVID inpatient utilisation, COVID vaccine and dyspnoea \cite{pfaff2022identifying}. Besides, the social determinants of health (SDOH) \cite{singu2020impact} are also severely impacting the susceptibility to COVID and Long COVID. Another study has focused on the risk factors associated with PASC \cite{Hill2022.08.15.22278603}, including common comorbidities and SDOH factors. As a result, middle age, several specific comorbidities and county level number of doctors are associated with Long COVID. However, there is limited attention to the vital measurements (e.g. oxygen saturation, heart rate and blood pressure) of Long COVID patients, and the pattern of the vital measurements of hospitalized patients are generally not understood. In this paper we study the vital measurements of the first 7 days since the hospitalization start date among COVID and Long COVID patients, and try to use information from vital measurements to predict Long COVID.   

Clinical prediction tasks including patient mortality and disease prediction are with much significance for early disease prevention and intervention. A recent study has found that the summary statistics of physiological time series (e.g. min, max, range, mean, standard deviation, skewness, kurtosis) \cite{Guo2020AnTasks} could play an important role in the prediction of length of hospital stay and patient mortality. Blood pressure is an important measure in clinical practice \cite{Chadachan2018UnderstandingPractice}, and the variability of blood pressure could imply health conditions. Summary statistics such as standard deviation, skewness and kurtosis could describe the variability of a variable, and thus it is of much interest to study the summary statistics and the time series of vital measurements. As for time series analysis of COVID, various studies have considered the number of cases as time series for forecasting \cite{maleki2020time}, but few studies have taken into consideration the patient level time series, such as the vital measurements.   
\section{Methodology}
\subsection{Principal component analysis}
Principal component analysis (PCA) is a linear dimension reduction technique in the mean-square error sense \cite{fodor2002survey}. Consider $n$ observations of a $p$-dimensional random variable. We denote the observation matrix by $\boldsymbol{X}_{p\times n}$. A linear dimension reduction technique seeks $k \leq p$ components of the new variable, being a linear combination of the original variables. 

Specifically for PCA, the new variables after the linear transformation are a few variables (the principal components) orthogonal to each other, and linear combinations of the original variables with largest variance. The first principal component has the largest variance, the second principal component is with the second largest variance and orthogonal to the first principal component, and so is this for all other principal components. Ideally, the first several principal components explain most of the variance. As a result of PCA, we transform a $p\times n$ data matrix into a $k\times n$ data matrix, and keep most of the information of the data set in a much lower dimension. It is helpful to understand the data set if different groups have obvious boundaries in a PCA plot. 
\subsection{Kolmogorov–Smirnov test}
In statistics, the Kolmogorov–Smirnov test (KS test) is a non-parametric test to compare a sample with a reference distribution (one-sample KS test) \cite{justel1997multivariate}, or to compare two samples (two-sample KS test) \cite{hodges1958significance}. The one-sample KS test focuses on the probability that the sample is drawn from the reference distribution, and the two-sample KS test focuses on the probability that the two samples are drawn from the same but unknown distribution. The KS test provides a practical tool to compare a sample to a known distribution, or compare samples with each other with unknown distribution. 
\subsection{Machine learning classification}
The goal of supervised learning is to approximate a function $f:X\rightarrow Y$ using a training set $S = \{x_i,y_i\}_{i=1}^N$ which could describe the relationship between $x$ and $y$. Specifically when $Y=\{0,1\}$, this is a binary classification problem. In the following we only consider supervised learning, which means the labels are all available \cite{melas2020mathematical}. The overall aim of a machine learning problem is to minimize the empirical risk: $\hat{f}= \argmin_{f\in\mathbb{F}}\frac{1}{N}\sum_{i=1}^NL(y_i,f(x_i))$. Commonly used classification techniques in supervised learning include logic-based algorithms (e.g. decision trees), perceptron-based techniques (e.g. neural networks), statistical learning algorithms (e.g. Naive Bayes and Bayesian networks), instance-based learning (e.g. $k$-nearest neighbor) and support vector machines \cite{kotsiantis2007supervised}. In this work, we use XGBoost, a scalable machine learning system for tree boosting \cite{chen2016xgboost}. This novel tree learning algorithm is suitable for handling sparse data with faster learning process.

\subsection{Time series methods}
\subsubsection{ARIMA models}
An autoregressive integrated moving average (ARIMA) model is a generalized version of autoregressive moving average (ARMA) model \cite{arima}. It is fitted to time series data to better understand the data or to predict future points in the series.
\subsubsection{Deep learning in time series classification}
Deep neural networks are widely used in time series classification tasks, such as Multi Layer Perceptron (MLP), Convolutional Neural Network (CNN), Echo State Network (ESN) \cite{ismail2019deep} and long short term memory Recurrent Neural Network (LSTM RNN) \cite{karim2017lstm}.

\section{Results and analysis}
As of completion of the paper, the N3C cohort \cite{n3ccohort} contains 5,274,332 patients with an active COVID-19 infection as indicated by a U07.1 code or a positive PCR or AG SARS-CoV-2 lab test, the first instance of which we use as their COVID-19 index date. Of these, 327,964 patients were hospitalized in the day prior to 16 days following the index SARS-CoV-2 PCR or AG lab result and a COVID-19 diagnosis of U07.1 was recorded in that same time period. Efforts have been made to harmonize the units and values from electronic health records in N3C \cite{bradwell2022harmonizing}, and in the following section we study vital measurements as shown in Table \rom{1}.
\begin{table}[htbp]
\caption{Vital measurements}

\begin{center}
\begin{tabular}{|p{3cm}|p{1.5cm}|p{1.5cm}|}
\hline
\textbf{Measured variable} & \textbf{Unit} & \textbf{Range}\\

\hline
SpO2 (oxygen saturation) & Percent & 0$\sim$100\\
\hline
Heart rate & Bpm & 0$\sim$500\\
\hline
Systolic blood pressure & mmHg & 0$\sim$400\\
\hline
Diastolic blood pressure & mmHg & 0$\sim$200\\
\hline

\end{tabular}
\label{tab1}
\end{center}
\end{table}
 
To study how the emergence of Long COVID may be predicted by pattern of vital measurements of the 327,964 hospitalized patients we needed to identify a Long COVID indicator. Due to limited documentation of Long COVID, we follow the work in \cite{pfaff2022coding}, where the Long COVID indicator is derived from a machine learning-based computable phenotype definition trained on cases where the U09.9 (Long COVID) diagnosis code was recorded. The computable phenotype assigns a likelihood score between 0 and 1, and in the following analysis, patients with computable phenotype values larger or equal to 0.75 are labeled as `Long COVID', and patients with computable phenotype values smaller or equal to 0.25 are labeled as `non Long COVID'. Because only a subset of patients are assigned values in these two ranges, we end up with a cohort of 85,196 patients who are all hospitalized around the time of the first known COVID infection and who we can assign a binary value for our Long COVID indicator. 

\subsection{Summary of the cohort}
In this cohort of 85,196 patients, the average and the median of the age at the time of first known COVID-19 infection are 57.5 and 60, white non-Hispanic patients consist of 55.4\% of the cohort, female patients consist of 51.75\% of the cohort, and 33.52\% patients of the cohort are Long COVID patients.
\begin{figure}
     \centering
     \begin{subfigure}[b]{0.15\textwidth}
         \centering
         \includegraphics[width=\textwidth]{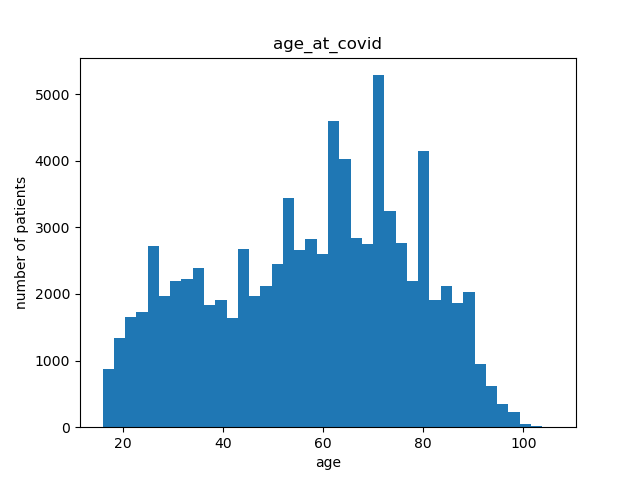}
         
         \label{fig:ageatcovid}
     \end{subfigure}
     \begin{subfigure}[b]{0.15\textwidth}
         \centering
         \includegraphics[width=\textwidth]{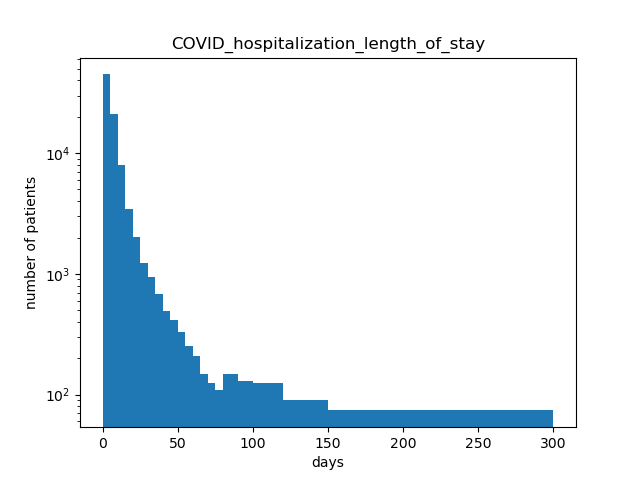}
         
         \label{fig:lengthinhospital}
     \end{subfigure}

        \caption{Age summary and length in hospital of the cohort}
        \label{fig:cohortinfo}
\end{figure}

\subsection{Vital measurements of the cohort of hospitalized patients}
To study the pattern of the vital measurements of Long COVID patients, we extract the vital measurements readings from the relevant COVID-associated hospitalizations.
The majority of the cohort have a short length of hospitalization, with an average of 8.1 days, and a median of 4 days. In order to reduce the impact of variable length of stay, we chose to focus on vitals collected during the first week of hospitalization. Before creating features from the vital measurements readings, we try to explore the distribution and the richness of the vital measurements readings. 

In the N3C cohort, records of patients are provided by anonymized institutions, represented by a variable `data partner id'. To comply with N3C policy, the data partner ids in Fig. 2 don't represent the real resources, and same data partner id belonging to different vital measurements might represent different resources. Table \rom{2} shows the amount of available readings of each vital measurement, and the number of patients that these readings belong to. Fig. 3 shows that the distribution of each vital measurement has a very high peak, and the distribution of oxygen saturation readings is less symmetric than the other 3 vital measurements. A KS test is performed on each vital measurement to test whether the readings are from a normal distribution, and the $p$-values of the KS test on each vital measurement are all 0, so we reject the null hypothesis that the readings are from a normal distribution.

With vital measurements readings in the first 7 days since the hospitalization start date, features describing overall conditions of the vital measurements are created as following:
\begin{enumerate}
  \item \textbf{Daily averages}: Respectively for each vital measurement, we take the daily average of readings of each patient for 7 days, and get a series with length 7.
  \item \textbf{Overall summary statistics features}: Respectively for each vital measurement, we calculate the summary statistics (e.g. min, max, median, quartiles, range, standard deviation, skewness and kurtosis) of all readings of each patient.
  \item \textbf{Daily variability features}: Respectively for each vital measurement, we first calculate the daily min, daily average, and daily max, and then calculate the variability measure (standard deviation, skewness, kurtosis) of the daily min, daily average and daily max of each patient. 
\end{enumerate}

\begin{table}[htbp]
\caption{Availability of vital measurements of the cohort}
\begin{center}
\begin{tabular}{|p{3cm}|p{1.2cm}|p{1cm}|p{1.6cm}|}
\hline
\textbf{Measurements} & {\textbf{Readings}} & {\textbf{Patients}} & {\textbf{Resources}}\\
\hline
Oxygen saturation & 6,959,178 & 44,479 & 59\\
\hline
Heart rate & 7,833,270 & 44,253 & 40 \\
\hline
Systolic blood pressure & 3,530,787 & 32,706 & 34 \\
\hline
Diastolic blood pressure & 3,840,918 & 35,628 & 34 \\
\hline
\end{tabular}
\label{tab7}
\end{center}
\end{table}

\begin{figure}
     \centering
     \begin{subfigure}[b]{0.1\textwidth}
         \centering
         \includegraphics[width=\textwidth]{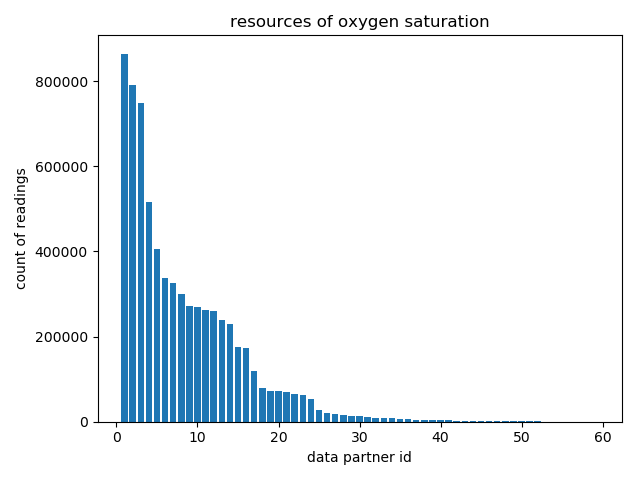}
         
         \label{fig:dpidos}
     \end{subfigure}
     \begin{subfigure}[b]{0.1\textwidth}
         \centering
         \includegraphics[width=\textwidth]{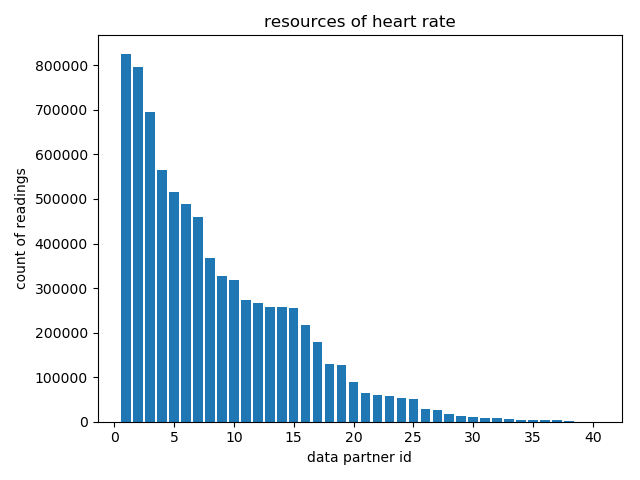}
         
         \label{fig:dpidhr}
     \end{subfigure}
     \begin{subfigure}[b]{0.1\textwidth}
         \centering
         \includegraphics[width=\textwidth]{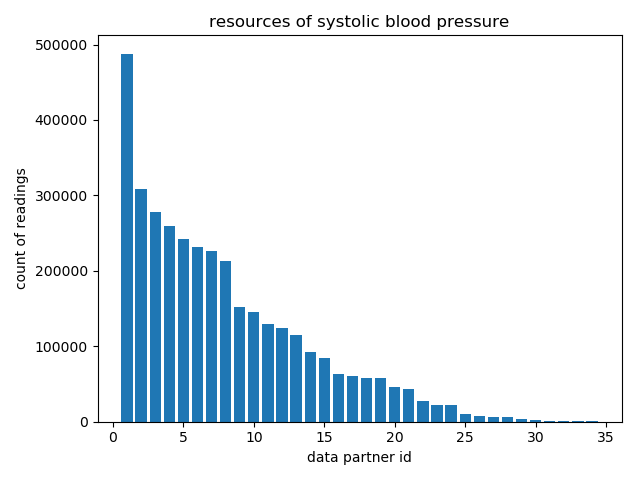}
         
         \label{fig:dpidsys}
     \end{subfigure}
     \begin{subfigure}[b]{0.1\textwidth}
         \centering
         \includegraphics[width=\textwidth]{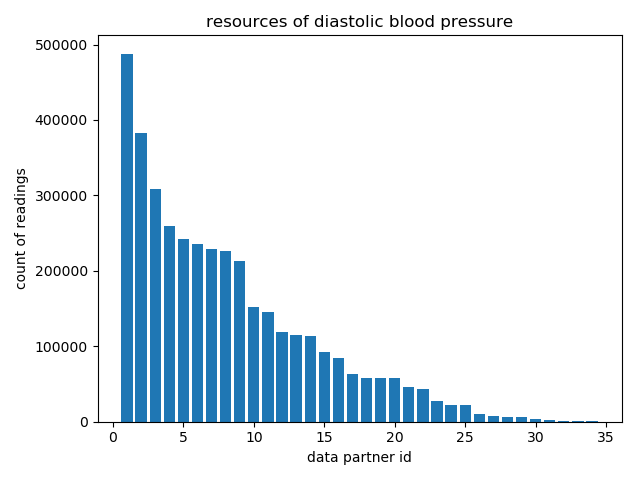}
         
         \label{fig:dpiddia}
     \end{subfigure}
        \caption{Resources of readings of the cohort}
        \label{fig:dpid}
\end{figure}

\begin{figure}
     \centering
     \begin{subfigure}[b]{0.2\textwidth}
         \centering
         \includegraphics[width=\textwidth]{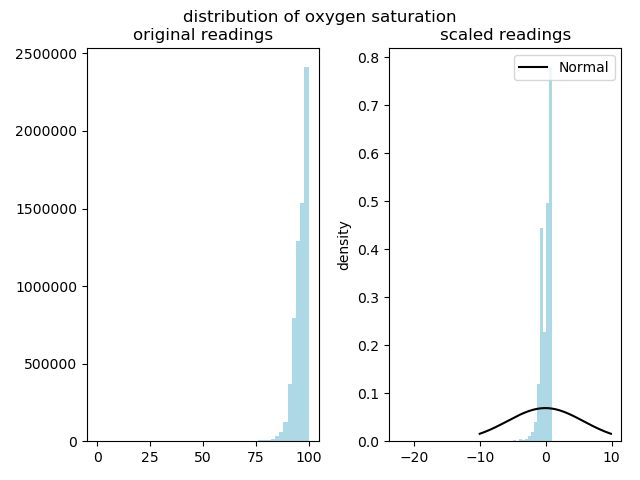}
         
         \label{fig:disos}
     \end{subfigure}
     \begin{subfigure}[b]{0.2\textwidth}
         \centering
         \includegraphics[width=\textwidth]{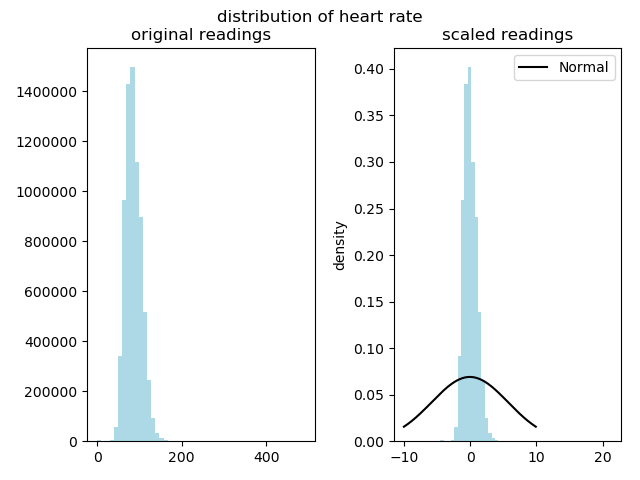}
         
         \label{fig:dishr}
     \end{subfigure}
     \begin{subfigure}[b]{0.2\textwidth}
         \centering
         \includegraphics[width=\textwidth]{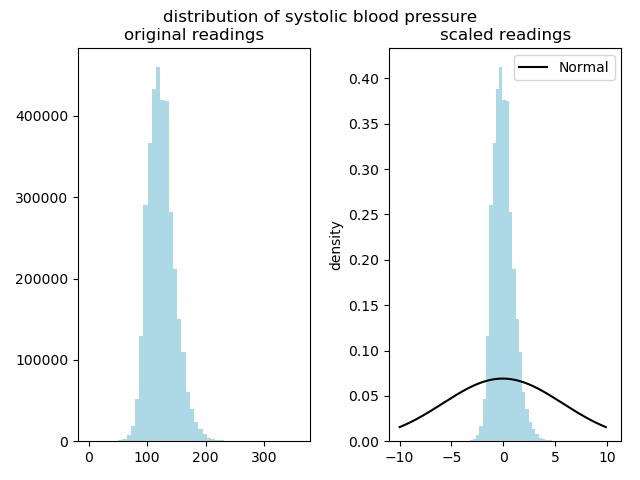}
         
         \label{fig:dissys}
     \end{subfigure}
     \begin{subfigure}[b]{0.2\textwidth}
         \centering
         \includegraphics[width=\textwidth]{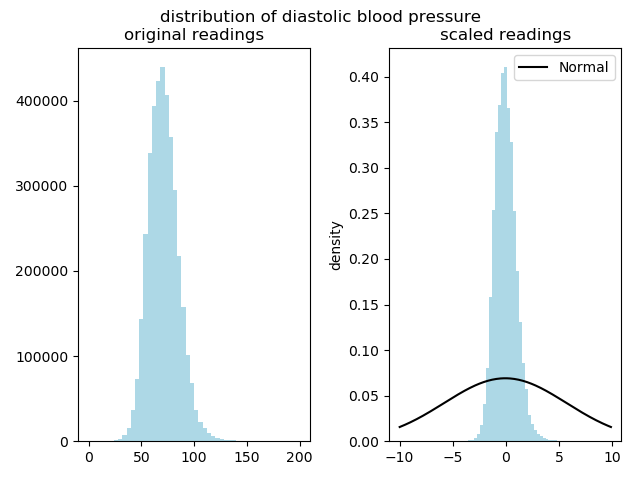}
         
         \label{fig:disdia}
     \end{subfigure}
        \caption{Distribution of readings of the cohort}
        \label{fig:dis}
\end{figure}

\subsection{Subcohort with rich data}
Despite the widely used electronic health record data, there are still only a relatively small portion of patients among the hospitalized patients with available vital measurements data, as Table \rom{2} shows. Thus, we select subcohorts from hospitalized patients with rich data for further numerical analysis of the vital measurements, especially the prediction of Long COVID using features created from vital measurements.
\subsubsection{Subcohort A}
The subcohort A consists of 16,468 patients with at least one reading of each of the four vital measurements (oxygen saturation, heart rate, systolic blood pressure, diastolic blood pressure). There are 8,385 female patients in the subcohort A, and among all patients in the subcohort A, 6,088 patients are labeled as `Long COVID', and the average and median of age of subcohort A are 56.5 and 59. Table \rom{3} shows the overall richness of vital measurements readings in 7 days of subcohort A.

\begin{table}[htbp]
\caption{Number of readings per patient of subcohort A}
\begin{center}
\begin{tabular}{|p{3cm}|p{2cm}|p{2cm}|}
\hline
\textbf{Measurements} & {\textbf{Average number of readings per patient}} & {\textbf{Median number of readings per patient}}\\
\hline
Oxygen saturation & 49.4 & 23 \\
\hline
Heart rate & 53.2 & 24 \\
\hline
Systolic blood pressure & 39.3 & 23 \\
\hline
Diastolic blood pressure & 38.8 & 23 \\
\hline
\end{tabular}
\label{tab8}
\end{center}
\end{table}

To get more information from the vital measurements, we focus on the feature set of vital measurements including daily averages, overall summary statistics and daily variability features of the four vital measurements (oxygen saturation, heart rate, systolic blood pressure, diastolic blood pressure). This feature set of vital measurements of subcohort A has 139 features, and a principal component analysis is performed on the vital measurements feature set. As Fig. 4 shows, red dots represent Long COVID patients, and blue dots represent non Long COVID patients in subcohort A, and the two principal components explain 19.74\% and 8.90\% of the total variance of the original feature set of vital measurements. It is desirable that there is an obvious boundary between these two groups. 

\begin{figure}[htbp]
\centerline
{\includegraphics[width=0.15\textwidth]{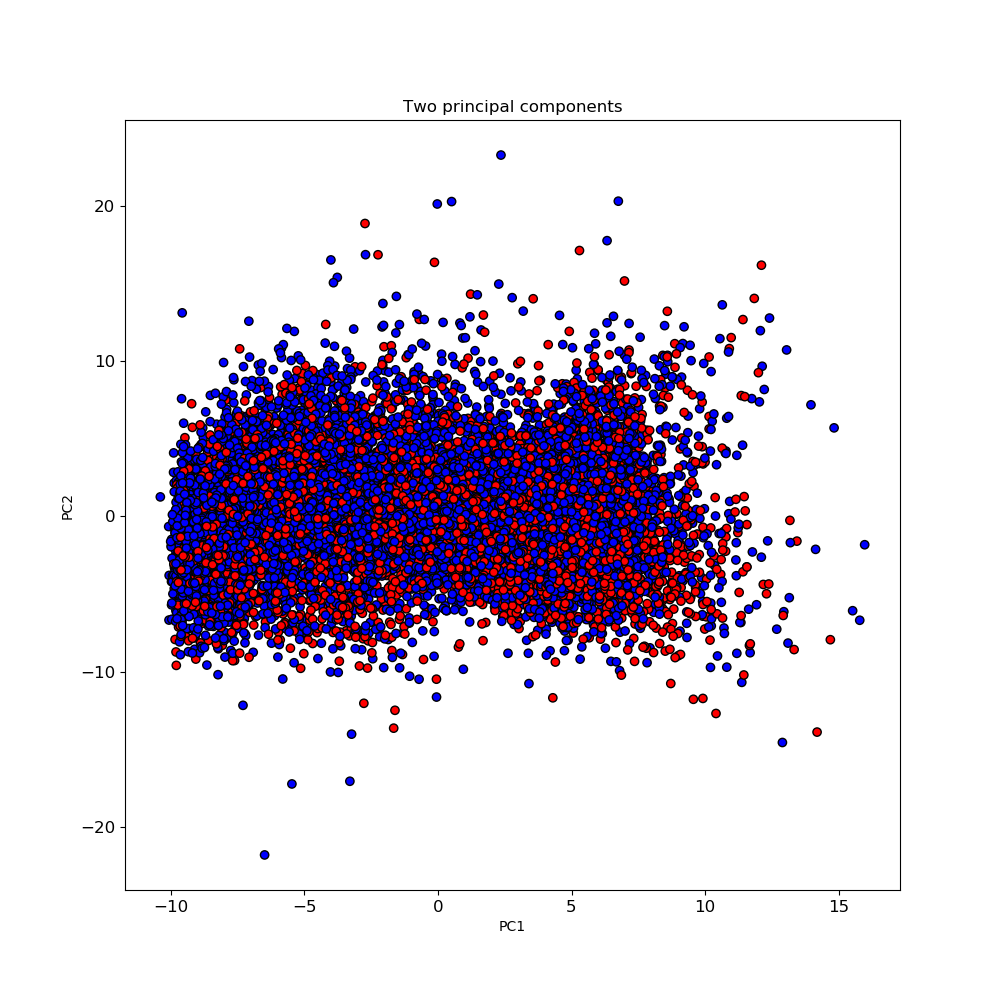}}
\caption{PCA analysis of the subcohort A}
\label{fig10}
\end{figure}

Besides, a KS test is performed on the feature set of vital measurements as well. The subcohort A is divided to two groups: the group of Long COVID patients and the group of non Long COVID patients. The KS test is to test whether the features under these two groups are from the same distribution. A feature with a large $p$-values means that we fail to reject the null hypothesis that the samples of this feature in two groups are from the same distribution. As a result, there are 118 features with $p$-value smaller than or equal to 0.01, and Table \rom{4} shows 4 features with a $p$-value larger than 0.1. 

\begin{table}[htbp]
\caption{KS test on subcohort A}
\begin{center}
\begin{tabular}{|p{4cm}|p{2cm}|}
\hline
\textbf{Feature} & \textbf{$p$-value} \\

\hline
third\_quartile\_dia\_bp\_7\_day & 0.1057 \\
\hline
mean\_dia\_bp\_1 & 0.2530 \\
\hline
mean\_dia\_bp\_2 & 0.2058\\
\hline
mean\_sys\_bp\_1 & 0.1621\\
\hline

\end{tabular}
\label{tab9}
\end{center}
\end{table}
As a result of the PCA and KS test, the two principal components could not separate the group of Long COVID and the group of non Long COVID, and most features have different distributions in these two groups. To better understand the role of vital measurements in the prediction of Long COVID, we train XGBoost models on the subcohort A respectively using 3 feature sets, where feature set 1 is referring to the risk factor analysis of Long COVID, but without age, gender and race and ethnicity of the cohort so as to avoid overlap with the feature set used to generate our Long COVID labels \cite{Hill2022.08.15.22278603}:
\begin{itemize}
  \item \textbf{Feature set 1 (SDOH and pre-COVID conditions)}: 51 features including social determinants of health (SDOH) and pre-COVID health conditions
  \item \textbf{Feature set 2 (vital measurements)}: 139 features created from vital measurements as mentioned above, including daily averages, overall summary statistics and daily variability features
  \item \textbf{Feature set 3 (SDOH, pre-COVID conditions and vital measurements)}: 190 features combining feature set 1 and feature set 2
\end{itemize}
With 5-fold cross validation, we evaluate performance of the models with the common metric, area under the ROC curve (AUC) and the F1 score (the harmonic mean of precision and recall) \cite{fawcett2006introduction}. Also, we calculate the permutation importance of each feature. As Table \rom{5} shows, in the subcohort A with rich vital measurements data, the vital measurement features perform better than the SDOH and pre-COVID conditions in the prediction of Long COVID, and adding the vital measurements features to SDOH and pre-COVID conditions could further improve the performance of the XGBoost model in prediction of Long COVID. In addition, we show the top 10 important features of each XGBoost model:
\begin{itemize}
  \item \textbf{Feature set 1 (SDOH and pre-COVID conditions)}: extended stay, chronic lung disease before COVID, Corticosteroid before COVID, dementia before COVID, hypertension before COVID, obesity before COVID, depression before COVID, percent insured 65 plus public, Corticosteroid during COVID hospitalization, long stay 
  \item \textbf{Feature set 2 (vital measurements)}: measurement duration of systolic blood pressure, measurement duration of diastolic blood pressure, observation per hour of diastolic blood pressure, measurement duration of oxygen saturation, minimum measurement time of heart rate since hospitalization, observation per hour of oxygen saturation, average of oxygen saturation, measurement duration of heart rate, maximum measurement time of heart rate since hospitalization, average of heart rate on the third day
  \item \textbf{Feature set 3 (SDOH, pre-COVID conditions and vital measurements)}: measurement duration of systolic blood pressure, chronic lung disease before COVID, extended stay, dementia before COVID, Corticosteroid before COVID, depression before COVID, Corticosteroid during COVID hospitalization, average of heart rate on the first day, measurement duration of diastolic blood pressure, measurement duration of oxygen saturation
\end{itemize}

\begin{table}[htbp]
\caption{XGBoost Results}
\begin{center}
\begin{tabular}{|p{2cm}|p{2cm}|p{2cm}|}
\hline
\textbf{Feature set} & \textbf{Mean AUC (5-fold CV)} & \textbf{Mean F1 score (5-fold CV)} \\
\hline
Feature set 1 & 0.729 $\pm$ 0.008 & 0.696 $\pm$ 0.007\\
\hline
Feature set 2 & 0.787 $\pm$ 0.007 & 0.717 $\pm$ 0.007\\
\hline
Feature set 3 & \textbf{0.822 $\pm$ 0.007} & \textbf{0.752 $\pm$ 0.005}\\
\hline

\end{tabular}
\label{tab10}
\end{center}
\end{table}
\subsubsection{Subcohort B}
The subcohort B consists of 5,304 patients with available daily averages of each of the four vital measurements (oxygen saturation, heart rate, systolic blood pressure, diastolic blood pressure) for first 7 consecutive days since the hospitalization start date. In other words, patients in subcohort B have 4 dimensional time series with length 7. There are 2,392 female patients in the subcohort B, and among all patients in the subcohort B, 2,237 patients are labeled as `Long COVID', and the average and median of age of subcohort B are 63.3 and 65.

To get an understanding of the trend of series, a regression model between average oxygen saturation value of 7th day and values of first 6 days is fitted on subcohort B. This regression model has a mean sqaured error of 2.615 and a $R^2$ score of 0.5691. Because of the low $R^2$ value, this regression model seems not relatively informative with a weak linear relationship, and the relationship inside the time series is still underneath the hood. Using the daily averages as features, common machine learning classification models are trained and evaluated as Table \rom{6} shows. Two neural network models are trained on subcohort B: one is using a fully convolutional neural network \cite{wang2017time} with 3 convolutional layers followed by the global average pooling process, and the other is using a layer of LSTM \cite{lstmlayer}. As a result, this CNN model has a test accuracy 0.6164 and a test loss 0.6673, and the LSTM model has a test accuracy of 0.5994 and a test loss of 0.6707. Fig. 5 shows the training loss and the validation loss with the number of epochs. 

\begin{table}[htbp]
\caption{Machine learning models on subcohort B}

\begin{center}
\begin{tabular}{|p{3cm}|p{3cm}|}
\hline
\textbf{Method} & \textbf{Mean AUC (5-fold CV)}\\
\hline
XGBoost & 0.614 $\pm$ 0.015\\
\hline
Logistic regression & 0.614 $\pm$ 0.015\\
\hline
Support vector machine & 0.617 $\pm$ 0.016\\
\hline
Naive Bayes & 0.608 $\pm$ 0.017\\
\hline
k-nearest neighbor (k=5) & 0.550 $\pm$ 0.014\\
\hline

\end{tabular}
\label{tab11}
\end{center}
\end{table}

\begin{figure}
     \centering
     \begin{subfigure}[b]{0.15\textwidth}
         \centering
         \includegraphics[width=\textwidth]{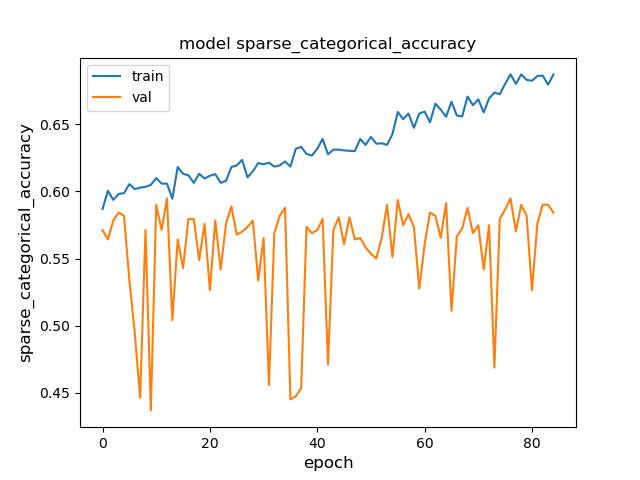}
         \caption{CNN}
         \label{fig:cnnsubb}
     \end{subfigure}
     \begin{subfigure}[b]{0.15\textwidth}
         \centering
         \includegraphics[width=\textwidth]{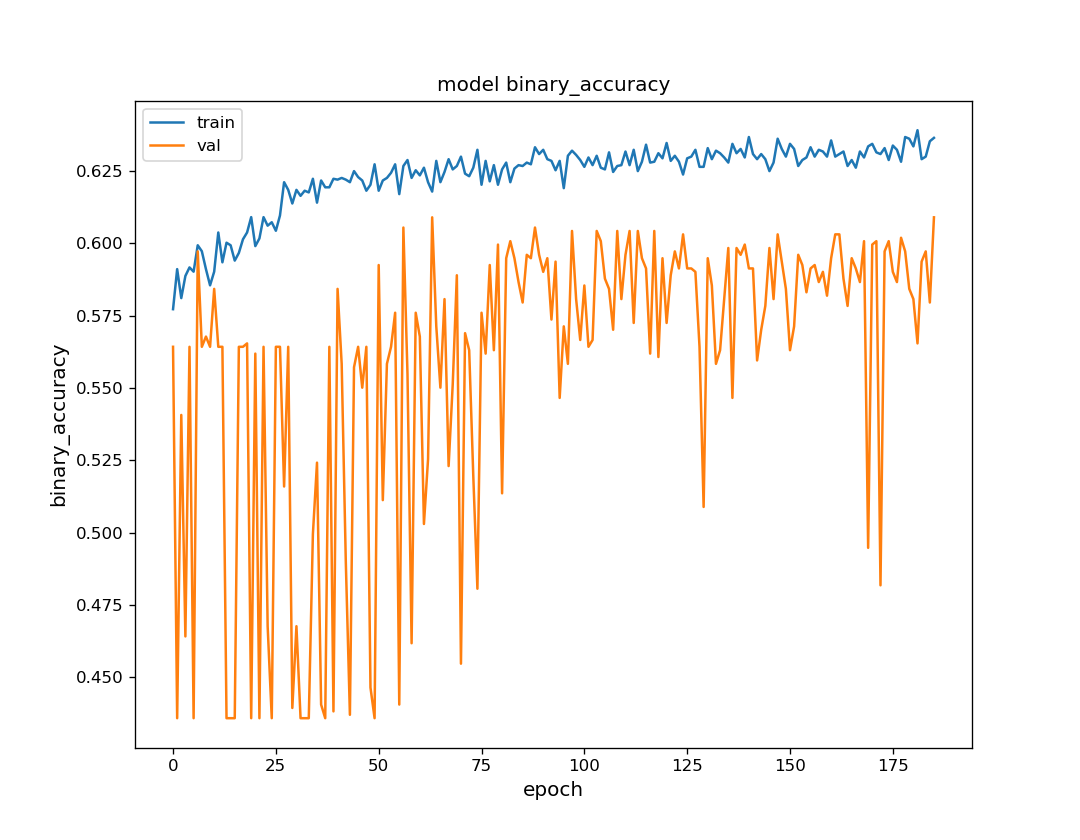}
         \caption{LSTM}
         \label{fig:lstmsubb}
     \end{subfigure}

        \caption{Results of CNN and LSTM models}
        \label{fig:nn}
\end{figure}
\subsection{Conclusion}
In the analysis of the cohort of Long COVID, the relationship between vital measurements and Long COVID has caught much attention. This cohort is balanced in age, gender and race and ethnicity, but not all of patients in this cohort have available data of vital measurements. Thus, two subcohorts with rich vital measurements data are created. Because the average and median of hospitalization length are short for most patients, we focus on the vital measurements in the first 7 days since the hospitalization start date, and various features are created from the vital measurements readings. The feature set of vital measurements has 139 features, including the daily averages, the summary statistics of the vital measurements and the daily variability features. PCA is a great method to reduce the dimension of the data set, as well as the size of the feature set. Using the SDOH and pre-COVID conditions, vital measurements and a combination of these two as feature sets, the XGBoost model is trained on the subcohort A and gives a prediction of Long COVID. As a result, the feature set of vital measurements outperforms the feature set of SDOH and pre-COVID conditions, and the combined feature set could further improve the accuracy of the prediction. The subcohort B contains time series of the vital measurements, making possible the use of neural networks in time series classification. CNN and LSTM are great tools of processing multidimensional time series. In summary, in subcohorts with rich vital measurements data, the vital measurements are informative and with significance in the analysis of Long COVID.

\section{Limitations and future work}
This paper gives a summary of explorations on the vital measurements of patients in the hospitalized Long COVID cohort, including the extracted features from vital measurements and the prediction of Long COVID using these features. However, since the Long COVID is still a new disease, there is only a small portion of patients with available vital measurements data in the whole N3C cohort, and more comprehensive data is desired. Besides, the biomedical meaning of the features created from vital measurements readings, such as the summary statistics of the time series, are generally unclear, so it is important to understand which features are medically significant. With an overall goal of predicting Long COVID with features created from vital measurements, we always need appropriate classification techniques that could predict Long COVID more accurately, including machine learning methods and deep neural networks. Methods of selecting models should be included in further analysis. While training models, the choice of hyperparameters would significantly influence the performance of models, and common methods of selecting hyperparameters include random search and grid search \cite{claesen2015hyperparameter}. These methods should be taken into consideration for better results. 
\section*{Acknowledgement}
The analyses described in this publication were conducted with data or tools accessed through the NCATS N3C Data Enclave covid.cd2h.org/enclave and supported by CD2H - The National COVID Cohort Collaborative (N3C) IDeA CTR Collaboration 3U24TR002306-04S2 NCATS U24 TR002306. This research was possible because of the patients whose information is included within the data from participating organizations (covid.cd2h.org/dtas) and the organizations and scientists (covid.cd2h.org/duas) who have contributed to the on-going development of this community resource \cite{haendel2021national}.

The content is solely the responsibility of the authors and does not necessarily represent the official views of the RECOVER Program, the National Institutes of Health or the N3C program.

Authorship has been determined according to ICMJE recommendations.

The N3C data transfer to NCATS is performed under a Johns Hopkins University Reliance Protocol \# IRB00249128 or individual site agreements with NIH. The N3C Data Enclave is managed under the authority of the NIH; information can be found at https://ncats.nih.gov/n3c/resources.

We gratefully acknowledge the following core contributors to N3C:
Adam B. Wilcox, Adam M. Lee, Alexis Graves, Alfred (Jerrod) Anzalone, Amin Manna, Amit Saha, Amy Olex, Andrea Zhou, Andrew E. Williams, Andrew Southerland, Andrew T. Girvin, Anita Walden, Anjali A. Sharathkumar, Benjamin Amor, Benjamin Bates, Brian Hendricks, Brijesh Patel, Caleb Alexander, Carolyn Bramante, Cavin Ward-Caviness, Charisse Madlock-Brown, Christine Suver, Christopher Chute, Christopher Dillon, Chunlei Wu, Clare Schmitt, Cliff Takemoto, Dan Housman, Davera Gabriel, David A. Eichmann, Diego Mazzotti, Don Brown, Eilis Boudreau, Elaine Hill, Elizabeth Zampino, Emily Carlson Marti, Emily R. Pfaff, Evan French, Farrukh M Koraishy, Federico Mariona, Fred Prior, George Sokos, Greg Martin, Harold Lehmann, Heidi Spratt, Hemalkumar Mehta, Hongfang Liu, Hythem Sidky, J.W. Awori Hayanga, Jami Pincavitch, Jaylyn Clark, Jeremy Richard Harper, Jessica Islam, Jin Ge, Joel Gagnier, Joel H. Saltz, Joel Saltz, Johanna Loomba, John Buse, Jomol Mathew, Joni L. Rutter, Julie A. McMurry, Justin Guinney, Justin Starren, Karen Crowley, Katie Rebecca Bradwell, Kellie M. Walters, Ken Wilkins, Kenneth R. Gersing, Kenrick Dwain Cato, Kimberly Murray, Kristin Kostka, Lavance Northington, Lee Allan Pyles, Leonie Misquitta, Lesley Cottrell, Lili Portilla, Mariam Deacy, Mark M. Bissell, Marshall Clark, Mary Emmett, Mary Morrison Saltz, Matvey B. Palchuk, Melissa A. Haendel, Meredith Adams, Meredith Temple-O'Connor, Michael G. Kurilla, Michele Morris, Nabeel Qureshi, Nasia Safdar, Nicole Garbarini, Noha Sharafeldin, Ofer Sadan, Patricia A. Francis, Penny Wung Burgoon, Peter Robinson, Philip R.O. Payne, Rafael Fuentes, Randeep Jawa, Rebecca Erwin-Cohen, Rena Patel, Richard A. Moffitt, Richard L. Zhu, Rishi Kamaleswaran, Robert Hurley, Robert T. Miller, Saiju Pyarajan, Sam G. Michael, Samuel Bozzette, Sandeep Mallipattu, Satyanarayana Vedula, Scott Chapman, Shawn T. O'Neil, Soko Setoguchi, Stephanie S. Hong, Steve Johnson, Tellen D. Bennett, Tiffany Callahan, Umit Topaloglu, Usman Sheikh, Valery Gordon, Vignesh Subbian, Warren A. Kibbe, Wenndy Hernandez, Will Beasley, Will Cooper, William Hillegass, Xiaohan Tanner Zhang. Details of contributions available at covid.cd2h.org/core-contributors

The following institutions whose data is released or pending:

Available: Advocate Health Care Network — UL1TR002389: The Institute for Translational Medicine (ITM) • Boston University Medical Campus — UL1TR001430: Boston University Clinical and Translational Science Institute • Brown University — U54GM115677: Advance Clinical Translational Research (Advance-CTR) • Carilion Clinic — UL1TR003015: iTHRIV Integrated Translational health Research Institute of Virginia • Charleston Area Medical Center — U54GM104942: West Virginia Clinical and Translational Science Institute (WVCTSI) • Children’s Hospital Colorado — UL1TR002535: Colorado Clinical and Translational Sciences Institute • Columbia University Irving Medical Center — UL1TR001873: Irving Institute for Clinical and Translational Research • Duke University — UL1TR002553: Duke Clinical and Translational Science Institute • George Washington Children’s Research Institute — UL1TR001876: Clinical and Translational Science Institute at Children’s National (CTSA-CN) • George Washington University — UL1TR001876: Clinical and Translational Science Institute at Children’s National (CTSA-CN) • Indiana University School of Medicine — UL1TR002529: Indiana Clinical and Translational Science Institute • Johns Hopkins University — UL1TR003098: Johns Hopkins Institute for Clinical and Translational Research • Loyola Medicine — Loyola University Medical Center • Loyola University Medical Center — UL1TR002389: The Institute for Translational Medicine (ITM) • Maine Medical Center — U54GM115516: Northern New England Clinical \& Translational Research (NNE-CTR) Network • Massachusetts General Brigham — UL1TR002541: Harvard Catalyst • Mayo Clinic Rochester — UL1TR002377: Mayo Clinic Center for Clinical and Translational Science (CCaTS) • Medical University of South Carolina — UL1TR001450: South Carolina Clinical \& Translational Research Institute (SCTR) • Montefiore Medical Center — UL1TR002556: Institute for Clinical and Translational Research at Einstein and Montefiore • Nemours — U54GM104941: Delaware CTR ACCEL Program • NorthShore University HealthSystem — UL1TR002389: The Institute for Translational Medicine (ITM) • Northwestern University at Chicago — UL1TR001422: Northwestern University Clinical and Translational Science Institute (NUCATS) • OCHIN — INV-018455: Bill and Melinda Gates Foundation grant to Sage Bionetworks • Oregon Health \& Science University — UL1TR002369: Oregon Clinical and Translational Research Institute • Penn State Health Milton S. Hershey Medical Center — UL1TR002014: Penn State Clinical and Translational Science Institute • Rush University Medical Center — UL1TR002389: The Institute for Translational Medicine (ITM) • Rutgers, The State University of New Jersey — UL1TR003017: New Jersey Alliance for Clinical and Translational Science • Stony Brook University — U24TR002306 • The Ohio State University — UL1TR002733: Center for Clinical and Translational Science • The State University of New York at Buffalo — UL1TR001412: Clinical and Translational Science Institute • The University of Chicago — UL1TR002389: The Institute for Translational Medicine (ITM) • The University of Iowa — UL1TR002537: Institute for Clinical and Translational Science • The University of Miami Leonard M. Miller School of Medicine — UL1TR002736: University of Miami Clinical and Translational Science Institute • The University of Michigan at Ann Arbor — UL1TR002240: Michigan Institute for Clinical and Health Research • The University of Texas Health Science Center at Houston — UL1TR003167: Center for Clinical and Translational Sciences (CCTS) • The University of Texas Medical Branch at Galveston — UL1TR001439: The Institute for Translational Sciences • The University of Utah — UL1TR002538: Uhealth Center for Clinical and Translational Science • Tufts Medical Center — UL1TR002544: Tufts Clinical and Translational Science Institute • Tulane University — UL1TR003096: Center for Clinical and Translational Science • University Medical Center New Orleans — U54GM104940: Louisiana Clinical and Translational Science (LA CaTS) Center • University of Alabama at Birmingham — UL1TR003096: Center for Clinical and Translational Science • University of Arkansas for Medical Sciences — UL1TR003107: UAMS Translational Research Institute • University of Cincinnati — UL1TR001425: Center for Clinical and Translational Science and Training • University of Colorado Denver, Anschutz Medical Campus — UL1TR002535: Colorado Clinical and Translational Sciences Institute • University of Illinois at Chicago — UL1TR002003: UIC Center for Clinical and Translational Science • University of Kansas Medical Center — UL1TR002366: Frontiers: University of Kansas Clinical and Translational Science Institute • University of Kentucky — UL1TR001998: UK Center for Clinical and Translational Science • University of Massachusetts Medical School Worcester — UL1TR001453: The UMass Center for Clinical and Translational Science (UMCCTS) • University of Minnesota — UL1TR002494: Clinical and Translational Science Institute • University of Mississippi Medical Center — U54GM115428: Mississippi Center for Clinical and Translational Research (CCTR) • University of Nebraska Medical Center — U54GM115458: Great Plains IDeA-Clinical \& Translational Research • University of North Carolina at Chapel Hill — UL1TR002489: North Carolina Translational and Clinical Science Institute • University of Oklahoma Health Sciences Center — U54GM104938: Oklahoma Clinical and Translational Science Institute (OCTSI) • University of Rochester — UL1TR002001: UR Clinical \& Translational Science Institute • University of Southern California — UL1TR001855: The Southern California Clinical and Translational Science Institute (SC CTSI) • University of Vermont — U54GM115516: Northern New England Clinical \& Translational Research (NNE-CTR) Network • University of Virginia — UL1TR003015: iTHRIV Integrated Translational health Research Institute of Virginia • University of Washington — UL1TR002319: Institute of Translational Health Sciences • University of Wisconsin-Madison — UL1TR002373: UW Institute for Clinical and Translational Research • Vanderbilt University Medical Center — UL1TR002243: Vanderbilt Institute for Clinical and Translational Research • Virginia Commonwealth University — UL1TR002649: C. Kenneth and Dianne Wright Center for Clinical and Translational Research • Wake Forest University Health Sciences — UL1TR001420: Wake Forest Clinical and Translational Science Institute • Washington University in St. Louis — UL1TR002345: Institute of Clinical and Translational Sciences • Weill Medical College of Cornell University — UL1TR002384: Weill Cornell Medicine Clinical and Translational Science Center • West Virginia University — U54GM104942: West Virginia Clinical and Translational Science Institute (WVCTSI)

Submitted: Icahn School of Medicine at Mount Sinai — UL1TR001433: ConduITS Institute for Translational Sciences • The University of Texas Health Science Center at Tyler — UL1TR003167: Center for Clinical and Translational Sciences (CCTS) • University of California, Davis — UL1TR001860: UCDavis Health Clinical and Translational Science Center • University of California, Irvine — UL1TR001414: The UC Irvine Institute for Clinical and Translational Science (ICTS) • University of California, Los Angeles — UL1TR001881: UCLA Clinical Translational Science Institute • University of California, San Diego — UL1TR001442: Altman Clinical and Translational Research Institute • University of California, San Francisco — UL1TR001872: UCSF Clinical and Translational Science Institute

Pending: Arkansas Children’s Hospital — UL1TR003107: UAMS Translational Research Institute • Baylor College of Medicine — None (Voluntary) • Children’s Hospital of Philadelphia — UL1TR001878: Institute for Translational Medicine and Therapeutics • Cincinnati Children’s Hospital Medical Center — UL1TR001425: Center for Clinical and Translational Science and Training • Emory University — UL1TR002378: Georgia Clinical and Translational Science Alliance • HonorHealth — None (Voluntary) • Loyola University Chicago — UL1TR002389: The Institute for Translational Medicine (ITM) • Medical College of Wisconsin — UL1TR001436: Clinical and Translational Science Institute of Southeast Wisconsin • MedStar Health Research Institute — UL1TR001409: The Georgetown-Howard Universities Center for Clinical and Translational Science (GHUCCTS) • MetroHealth — None (Voluntary) • Montana State University — U54GM115371: American Indian/Alaska Native CTR • NYU Langone Medical Center — UL1TR001445: Langone Health’s Clinical and Translational Science Institute • Ochsner Medical Center — U54GM104940: Louisiana Clinical and Translational Science (LA CaTS) Center • Regenstrief Institute — UL1TR002529: Indiana Clinical and Translational Science Institute • Sanford Research — None (Voluntary) • Stanford University — UL1TR003142: Spectrum: The Stanford Center for Clinical and Translational Research and Education • The Rockefeller University — UL1TR001866: Center for Clinical and Translational Science • The Scripps Research Institute — UL1TR002550: Scripps Research Translational Institute • University of Florida — UL1TR001427: UF Clinical and Translational Science Institute • University of New Mexico Health Sciences Center — UL1TR001449: University of New Mexico Clinical and Translational Science Center • University of Texas Health Science Center at San Antonio — UL1TR002645: Institute for Integration of Medicine and Science • Yale New Haven Hospital — UL1TR001863: Yale Center for Clinical Investigation

\section*{Funding Statement}
This research is funded by the National Institutes of Health (NIH) Agreement OT2HL161847-01 and the NIH award UL1TR003015 for the integrated Translational Health Research Institute of Virginia.

\bibliographystyle{IEEEtran}
\bibliography{references}
\end{document}